\documentclass[journal]{IEEEtran}
\usepackage{amsmath,amsfonts}
\usepackage{algorithmic}
\usepackage{algorithm}
\usepackage{array}
\usepackage[caption=false,font=normalsize,labelfont=sf,textfont=sf]{subfig}
\usepackage{textcomp}
\usepackage{stfloats}
\usepackage{url}
\usepackage{verbatim}
\usepackage{graphicx}
\usepackage{cite}
\usepackage{amssymb}
\usepackage{mathtools}
\usepackage{csquotes}
\usepackage[final]{microtype}
\usepackage{xurl}
\usepackage{hyperref}
\hypersetup{hidelinks}
\usepackage{xcolor}
\newcommand{\rev}[1]{#1}

\usepackage{tikz}
\usetikzlibrary{arrows.meta,calc,angles,quotes}

\begin{document}

\title{Complex-Vector Power and Cross-Phase Unbalance in Three-Phase Systems}

\author{Juan Carlos Bravo-Rodr\'iguez, Juan Carlos del-Pino-L\'opez, Francisco Casado-Machado
% <-this % stops a space
\thanks{Juan Carlos Bravo-Rodr\'iguez, Juan Carlos del-Pino-L\'opez, Francisco Casado-Machado are with the Department of Electrical Engineering, University of Seville, Seville, Spain (e-mail: carlos\_bravo@us.es).}%        
%\thanks{This paper was produced by the IEEE Publication Technology Group. They are in Piscataway, NJ.}% <-this % stops a space
%\thanks{Manuscript received XXXX XX, 2026; revised XXXXXXXX XX, 2026.}
}

%\markboth{Journal of \LaTeX\ Class Files,~Vol.~XX, No.~X, March~2026}%


\maketitle

\begin{abstract}
Unbalanced three-phase systems still lack a compact phasor-domain representation of power that makes the joint cross-phase voltage--current structure explicit while remaining consistent with established apparent-power definitions. This paper addresses that point through a complex-vector power (CVP) formulation for sinusoidal steady-state operation. \rev{The proposed CVP is an element of a direct-sum space whose constituents are the conventional complex-power scalar and a cross-phase unbalance (CPU) vector obtained from the cross product of the voltage and current phasor vectors. The scalar constituent retains the usual active and reactive powers, whereas the CPU vector characterizes their antisymmetric cross-phase relation.} Apparent power is thereby separated into intraphase and cross-phase contributions, and its norm is preserved under the power-invariant Fortescue transformation. The formulation is extended to three-phase four-wire systems by introducing equivalent coordinates that preserve the effective apparent-power norm for the chosen voltage reference. Only standard complex numbers and matrices are required. \rev{Numerical examples identify operating conditions in which a non-negligible part of the apparent-power structure is associated with cross-phase unbalance and is not displayed by the active- and reactive-power pair alone. A proof-of-concept shunt-compensation study further illustrates how the CPU descriptor can be incorporated into reproducible current references. The proposed formulation therefore provides a compact phasor-based description of unbalance that complements established scalar apparent-power representations.
}

\end{abstract}

\begin{IEEEkeywords}
three-phase unbalance, symmetrical components, apparent power, complex-vector power, phasor-based methods, three-phase four-wire systems, power quality
\end{IEEEkeywords}

\section{Introduction}

Three-phase systems are meant to operate under balanced conditions, but real
networks rarely do. Unequal single-phase loading, conductor asymmetries, and
faults disturb the voltage and current sets and alter the way apparent power and
power factor must be interpreted. In those situations, the conventional active
and reactive power pair does not fully describe the structure of the electrical
regime.

The method of symmetrical components, introduced by
Fortescue~\cite{Fortescue1918,Fortescue1920}, remains the standard tool for
analyzing unbalanced three-phase phasors. By splitting the system into
positive-, negative-, and zero-sequence subsystems, it provides an effective
description of asymmetry. This work deliberately focuses on sinusoidal
steady-state unbalance. This is not intended as a simplification of the general
power-quality problem, but as an isolation of one of its fundamental
components: the description of apparent power when voltage and current phasors
are asymmetric while waveform distortion is absent. Establishing a clear
phasor-based formulation for this case is a necessary step before additional
effects such as harmonic distortion or time-varying behavior are introduced.

Several established theories deal with apparent power under imbalance.
Classical effective-apparent-power formulations such as the
Fryze--Buchholz--Depenbrock (FBD) method~\cite{Depenbrock1993}, the Currents'
Physical Components (CPC) framework~\cite{Czarnecki2006}, and the practical
line of work developed by
Emanuel~\cite{Emanuel1993,Emanuel1998,Emanuel1999} and later reflected in IEEE
Std.~1459--2010~\cite{IEEE1459} provide scalar measures and useful component
decompositions. Other contributions have explored vector \rev{and reference-frame} representations in
the sinusoidal regime~\cite{Leon2007,Diez2016,LeonMartinez2018,Blasco2020,CasadoMachado2023},
while broader algebraic approaches have extended the discussion to
nonsinusoidal and multiphase settings~\cite{LevAri2009,Montoya2025}.

\rev{Beyond the description of unbalanced operating conditions, their mitigation is commonly addressed through sequence-based control, active filtering, and multifunctional inverter operation~\cite{Vijay2020,Castilla2025,Duarte2024}. These approaches require suitable quantities to characterize the operating state and define compensation objectives. Most conventional indices describe voltage or current asymmetry separately, whereas a descriptor based on their joint relation may provide complementary information on the associated apparent-power structure.
}

\rev{Against this background, the reviewed power representations and
unbalance indicators do not simultaneously provide, within the conventional
sinusoidal phasor framework,} a compact quantity built directly from terminal
phasors that retains the classical complex-power interpretation, makes the
antisymmetric cross-phase voltage--current relation explicit, and preserves
the apparent-power norm under the power-invariant Fortescue transformation.
\rev{Such a representation could complement conventional sequence-based
indices by providing additional apparent-power information for the assessment
of unbalanced operating conditions and the formulation of compensation
objectives.} This combined
representation is the main focus of the present work.

\rev{To address this point, the paper introduces the Complex-Vector Power
(CVP), defined as a structured apparent-power representation composed of the
conventional complex-power scalar and a cross-phase unbalance (CPU) vector obtained
from the cross product of voltage and current phasors.} The scalar term retains the usual active and reactive powers, whereas the CPU vector is interpreted not as a new physical
form of power, but as a structural descriptor of the antisymmetric cross-phase relation that
becomes explicit when the phase-domain voltage--current relation is considered
as a whole.

The formulation remains fully within the standard phasor framework used in
power-system analysis. No advanced algebraic machinery is required: the
construction relies only on complex arithmetic and elementary matrix
operations. This makes the approach directly compatible with conventional
engineering practice while providing explicit information about cross-phase
asymmetry that is not displayed by scalar apparent-power representations
alone.

\rev{This compatibility also clarifies the relationship between the CVP and
the underlying network-analysis method. General-purpose circuit and
distribution-system solvers, including companion-circuit
approaches~\cite{Dommel1969} and OpenDSS-based
approaches~\cite{Dugan2011}, determine the electrical operating state
from the network topology and component models, whereas physical
measurements may provide the corresponding terminal quantities directly. The CVP is not a circuit-analysis solver and is not intended to replace network simulation. It is a phasor-domain power representation evaluated from the terminal voltage and current phasors once they have been obtained. Accordingly, effects such as mutual coupling may be incorporated into the underlying network model without modifying the algebraic definition of the CVP, although they affect the resulting phasors and, consequently, the numerical values of the CVP quantities.}

\rev{The scope of the present formulation is nevertheless restricted to unbalanced three-phase systems under sinusoidal steady-state conditions. Within this domain, the cross-phase term can be represented by the ordinary three-dimensional cross product. Extensions to higher phase orders would require a more general antisymmetric product and are therefore left outside the scope of this paper.}

\rev{Within this three-phase setting,} the CVP norm is preserved under the
power-invariant Fortescue transformation, providing a consistent link between
phase and sequence representations. \rev{The apparent-power norm is thereby
obtained from the orthogonal combination of an intraphase contribution,
associated with the dot-product term, and a cross-phase contribution associated
with the CPU vector.} The formulation is also extended to three-phase
four-wire systems by introducing equivalent coordinates that preserve the
effective apparent-power norm associated with the selected voltage reference,
in agreement with IEEE~1459 in the sinusoidal steady-state regime.

The main contributions of this paper are therefore fourfold. First, it defines a Complex-Vector Power that combines the conventional complex-power scalar and a cross-phase unbalance vector in a structured apparent-power representation. Second, it characterizes the CPU vector as a descriptor of the mutual voltage--current relation and establishes the corresponding norm decomposition. Third, it proves preservation of the CVP norm under the power-invariant Fortescue transformation. Finally, it extends the construction to three-phase four-wire systems through equivalent coordinates compatible with effective apparent power. \rev{The practical use of the CPU descriptor is subsequently illustrated
through a limited proof-of-concept assessment showing selective CPU
reduction and continuous $Q$--CPU allocation under a prescribed power
factor.}

These contributions are developed progressively in the remainder of the
paper. Section~II introduces the CVP and formulates it for three-phase systems.
Section~III extends the approach to four-wire configurations. Section~IV
derives an instantaneous representation of the cross-phase term. Section~V presents numerical examples \rev{and the limited compensation assessment,}
after which the main implications of the formulation are discussed and
summarized.

\section{CVP in Three-Phase Systems}

\rev{The notation used in this paper is as follows. Nonbold italic symbols denote scalar quantities, an overbar identifies complex scalar quantities such as phasors, impedances, and vector components, and bold symbols denote vectors or matrices. The superscript ${}^{*}$ denotes complex conjugation, applied elementwise to vectors and matrices, whereas ${}^{T}$ denotes transpose. The superscript ${}^{\pm0}$ denotes a vector expressed in symmetrical-component coordinates, whereas the superscripts ${}^{+}$, ${}^{-}$, and ${}^{0}$ identify its positive-, negative-, and zero-sequence components, respectively.}

\subsection{Definition}

In sinusoidal steady-state operation, voltages and currents in a three-phase system are represented by the phasor vectors
        \begin{equation}
\mathbf{V}=
\begin{bmatrix}
\bar V_{1} \\[2pt]
\bar V_{2} \\[2pt]
\bar V_{3}
\end{bmatrix},
\qquad
\mathbf{I}=
\begin{bmatrix}
\bar I_{1} \\[2pt]
\bar I_{2} \\[2pt]
\bar I_{3}
\end{bmatrix},
\end{equation}
which, in this section, are assumed to be defined with respect to an arbitrary but fixed voltage reference; a specific choice is introduced only when the CVP is aligned with a particular effective apparent-power definition, as discussed in Section~III.

\rev{This construction is illustrated in Fig.~\ref{fig:complex_vector_construction}.}

\begin{figure}[ht]
\centering
\makebox[\columnwidth][l]{%
\begin{tikzpicture}[
>=Latex,
scale=0.75,
transform shape,
axis/.style={->, gray!75},
Vphasor/.style={->, thick, black},
Iphasor/.style={->, thick, black, dashed},
proj/.style={dashed, gray!65, thick}
]

% =====================================================
% CONVENTIONAL PHASOR REPRESENTATION
% =====================================================

\begin{scope}[xshift=-0.15cm]

\node[font=\bfseries\small] at (0,3.65) {Conventional phasor representation};

\draw[axis] (-2.45,0) -- (2.55,0) node[right] {$\Re$};
\draw[axis] (0,-2.25) -- (0,2.35) node[above] {$\Im$};

% Voltage phasors
\draw[Vphasor] (0,0) -- ({93.07/35*cos(-3.95)},{93.07/35*sin(-3.95)})
node[below] {$\bar V_{1}$};

\draw[Vphasor] (0,0) -- ({91.19/35*cos(-123.71)},{91.19/35*sin(-123.71)})
node[left] {$\bar V_{2}$};

\draw[Vphasor] (0,0) -- ({91.83/35*cos(118.93)},{91.83/35*sin(118.93)})
node[left] {$\bar V_{3}$};

% Current phasors
\draw[Iphasor] (0,0) -- ({2.44*0.75*cos(-35.28)},{2.44*0.75*sin(-35.28)})
node[below right] {$\bar I_{1}$};

\draw[Iphasor] (0,0) -- ({4*0.75*cos(-161.14)},{4*0.75*sin(-161.14)})
node[below right] {$\bar I_{2}$};

\draw[Iphasor] (0,0) -- ({2.89*0.75*cos(65.15)},{2.89*0.75*sin(65.15)})
node[above right] {$\bar I_{3}$};

\node[align=center, font=\footnotesize] at (0,-3.15)
{Phase quantities are represented\\
as complex phasors in one plane};

\end{scope}

% =====================================================
% ORTHOGONAL COMPLEX-VECTOR REPRESENTATION
% =====================================================

\begin{scope}[xshift=5.4cm]

\node[font=\bfseries\small] at (0.3,3.65) {Orthogonal complex-vector representation};

\coordinate (O) at (0,0);
\coordinate (Ea) at (3.25,-0.66);
\coordinate (Eb) at (-2.15,-1.00);
\coordinate (Ec) at (0,2.95);

\draw[axis] (O) -- (Ea) node[below] {$\mathbf {e}_2$};
\draw[axis] (O) -- (Eb) node[below] {$\mathbf {e}_1$};
\draw[axis] (O) -- (Ec) node[right] {$\mathbf {e}_3$};

% Component endpoints
\coordinate (Va) at (2.25,-0.46);
\coordinate (Vb) at (-1.45,-0.67);
\coordinate (Vc) at (0,2.25);

\coordinate (Ia) at (2.55,-0.52);
\coordinate (Ib) at (-0.92,-0.42);
\coordinate (Ic) at (0,1.60);

% Resultant points
\coordinate (Vab) at ($(Va)+(Vb)$);
\coordinate (Vres) at ($(Va)+(Vb)+(Vc)$);

\coordinate (Iab) at ($(Ia)+(Ib)$);
\coordinate (Ires) at ($(Ia)+(Ib)+(Ic)$);

% Projection lines
\draw[proj] (Va) -- (Vab);
\draw[proj] (Vb) -- (Vab);
\draw[proj] (Vab) -- (Vres);
\draw[proj] (Vc) -- (Vres);

\draw[proj] (Ia) -- (Iab);
\draw[proj] (Ib) -- (Iab);
\draw[proj] (Iab) -- (Ires);
\draw[proj] (Ic) -- (Ires);

% Folded voltage components
\draw[->, black!55, thick] (O) -- (Va)
node[midway, above, xshift=18pt, yshift=-4pt] {$\bar V_{2}$};

\draw[->, black!55, thick] (O) -- (Vb)
node[midway, left, xshift=-10pt, yshift=2pt] {$\bar V_{1}$};

\draw[->, black!55, thick] (O) -- (Vc)
node[left] {$\bar V_{3}$};

% Folded current components
\draw[->, black!55, thick, dashed] (O) -- (Ia)
node[above right, xshift=-8pt] {$\bar I_{2}$};

\draw[->, black!55, thick, dashed] (O) -- (Ib)
node[midway, above left, xshift=-1pt, yshift=-2pt] {$\bar I_{1}$};

\draw[->, black!55, thick, dashed] (O) -- (Ic)
node[left] {$\bar I_{3}$};

% Main 3D complex vectors
\draw[->, line width=1.4pt, black] (O) -- (Vres)
node[above left, xshift=12pt, fill=white, inner sep=1pt] {$\mathbf V$};

\draw[->, line width=1.4pt, black, dashed] (O) -- (Ires)
node[right, fill=white, inner sep=1pt] {$\mathbf I$};

% Angle
\draw[gray!80, thick] (0.56,0.22) arc[start angle=32,end angle=55,radius=0.70];
\node[font=\small, fill=white, inner sep=1pt] at (0.64,0.46) {$\theta$};

\node[align=center, font=\footnotesize] at (0.3,-3.15)
{Each phasor becomes a complex\\
component of an orthogonal vector};

\end{scope}

\end{tikzpicture}%
}

\caption{\rev{Construction of the orthogonal complex-vector representation adopted in the proposed CVP formulation. Conventional phase phasors are interpreted as complex components of the voltage and current vectors.}}
\label{fig:complex_vector_construction}
\end{figure}

The classical complex power follows from the dot product
\begin{equation}
\mathbf{V}\cdot \mathbf{I}^{*}
=\sum_{k=1}^{3} \bar V_{k}\bar I_{k}^{*}
=P+jQ,
\end{equation}
which aggregates intraphase active and reactive contributions. However, the dot product does not capture antisymmetric phase-to-phase relations between voltage and current phasors that arise when the system departs from balanced operation.

These relations are captured by the cross product of the phasors, which defines a cross-phase unbalance (CPU) vector
\begin{equation}
\mathbf{D}=\mathbf{V}\times\mathbf{I},
\end{equation}
whose components are
\begin{align}
\bar D_{1}&=\bar V_{2}\bar I_{3}-\bar V_{3}\bar I_{2}, \nonumber\\
\bar D_{2}&=\bar V_{3}\bar I_{1}-\bar V_{1}\bar I_{3}, \nonumber\\
\bar D_{3}&=\bar V_{1}\bar I_{2}-\bar V_{2}\bar I_{1}.
\end{align}

\rev{The notation $\mathbf{D}$ is used here only to denote the CPU vector and should not be confused with distortion-power terms associated with nonsinusoidal power theories.}

\rev{The \emph{Complex-Vector Power} (CVP) is defined as the structured apparent-power representation whose constituents are the conventional complex-power scalar and the cross-phase unbalance vector introduced above,}
\begin{equation}
\mathbf{S}\triangleq
\left(
\mathbf{V}\cdot\mathbf{I}^{*},
\mathbf{V}\times\mathbf{I}
\right)
=
(P+jQ)\oplus\mathbf{D}
\in\mathbb{C}\oplus\mathbb{C}^{3},
\label{eq:CVP}
\end{equation}
\rev{where $\oplus$ denotes the direct-sum composition of the scalar constituent $P+jQ\in\mathbb{C}$ and the vector constituent $\mathbf{D}\in\mathbb{C}^{3}$. Thus, it does not denote arithmetic addition between quantities belonging to different mathematical spaces.}

\rev{The metric foundation of this structured representation is provided by the complex Lagrange identity, whose derivation is given in Appendix~A. Since $\mathbf{V}\cdot\mathbf{I}^{*}=P+jQ$ and $\mathbf{V}\times\mathbf{I}=\mathbf{D}$, this identity yields}
\begin{equation}
\|\mathbf{S}\|^{2}
\triangleq
\|\mathbf{V}\|^{2}\|\mathbf{I}\|^{2}
=
P^{2}+Q^{2}+\|\mathbf{D}\|^{2}.
\label{eq:CVP_norm}
\end{equation}
\rev{Therefore, the complex Lagrange identity establishes an orthogonal decomposition, in the direct-sum metric, between the intraphase constituent $P+jQ$ and the cross-phase constituent $\mathbf{D}$, while the resulting norm defines the apparent-power magnitude associated with the CVP.}

\rev{For a balanced phase-decoupled load with equal phase admittances, the terminal voltage and current vectors are complex-proportional,
\begin{equation}
\mathbf{I}=\alpha\,\mathbf{V},
\qquad
\alpha\in\mathbb{C},
\end{equation}
so that the cross product vanishes, $\mathbf{D}=\mathbf{0}$, and the classical complex power $P+jQ$ is the only possible nonzero constituent of the CVP. More generally, the condition $\mathbf{D}=\mathbf{0}$ expresses complex proportionality between the voltage and current vectors, but does not, by itself, imply that either phasor set is balanced. Voltage and current asymmetry concerns each set individually, whereas the CPU vector characterizes their mutual cross-phase relation.}

\rev{The orthogonal decomposition established by the complex Lagrange identity admits the schematic geometric interpretation illustrated in Fig.~\ref{fig:CVP_geometry}.}

\begin{figure}[ht]
\centering
\begin{tikzpicture}[
>=Latex,
scale=1.10,
transform shape,
axis/.style={->, gray!75},
proj/.style={dashed, gray!65, thick},
vecS/.style={->, very thick, black},
vecPQ/.style={->, thick, black},
vecD/.style={->, thick, black}
]

% Origin
\coordinate (O) at (0,0);

% Oblique 3D projection basis
\coordinate (Qaxis) at (3.75,-0.76);
\coordinate (Paxis) at (-2.45,-1.14);
\coordinate (Daxis) at (0,2.65);

% Numerical values, visually scaled:
% P=648.66, Q=542.72, ||D_e||=228.40
% Q is shown slightly compressed for readability.
\coordinate (Qp) at (2.60,-0.53);
\coordinate (Pp) at (-1.71,-0.80);
\coordinate (Dp) at (0,1.92);

% Construction points
\coordinate (Sp) at ($(Pp)+(Qp)$);
\coordinate (Stot) at ($(Sp)+(Dp)$);

% Axes
\draw[axis] (O) -- (Qaxis) node[right] {$e_Q$};
\draw[axis] (O) -- (Paxis) node[left] {$e_P$};
\draw[axis] (O) -- (Daxis) node[above] {$e_D$};

% Projections in the P-Q-D space
\draw[proj] (Pp) -- (Sp);
\draw[proj] (Qp) -- (Sp);
\draw[proj] (Sp) -- (Stot);
\draw[proj] (Dp) -- (Stot);

% P and Q scalar components
\draw[->, thick, black!70] (O) -- (Pp)
node[midway, above left] {$P$};

\draw[->, thick, black!70] (O) -- (Qp)
node[midway,above, right, xshift=20pt, yshift=3pt] {$Q$};

% Classical scalar complex power projection
\draw[vecPQ] (O) -- (Sp)
node[below , xshift=-3pt, yshift=3pt] {$|P+jQ|$};

% D norm component
\draw[vecD] (Sp) -- (Stot)
node[above, right, xshift=0pt, yshift=-10pt] {$\|\mathbf D\|$};

% Total apparent CVP norm
\draw[vecS] (O) -- (Stot)
node[above left, xshift=-5pt, yshift=-3pt,inner sep=1pt] {$\|\mathbf S\|$};

% Angles, shown schematically
\draw[black!70, thick] (-0.50,-0.24) arc[start angle=-140,end angle=-63,radius=0.62];
\node[black!70] at (-0.250,-0.66) {$\varphi$};

\draw[black!70, thick] (0.38,-0.58) arc[start angle=-50,end angle=40,radius=0.62];
\node[black!70] at (0.72,-0.32) {$\theta$};

\end{tikzpicture}
\caption{\rev{Schematic norm-space representation of the CVP decomposition. The $P$--$Q$ plane provides the conventional geometric representation of the complex scalar $P+jQ$, whereas the third coordinate represents only the magnitude $\|\mathbf D\|$ of the CPU-vector constituent. The diagram illustrates the CVP norm relation and does not identify $\mathbf D$ with a scalar physical-power axis. Component lengths are shown schematically and are not drawn to a common numerical scale.}}

\label{fig:CVP_geometry}
\end{figure}

\rev{In the schematic norm-space representation of
Fig.~\ref{fig:CVP_geometry}, the angle $\varphi$ retains its conventional
meaning within the scalar $P$--$Q$ plane and is defined by}
\begin{equation}
\tan\varphi
=
\frac{Q}{P},
\qquad
-\pi<\varphi\leq\pi.
\label{eq:phi_angle}
\end{equation}
\rev{The appropriate quadrant is determined from the signs of $P$ and $Q$.}

\rev{Consistently with the norm decomposition established by the complex
Lagrange identity, the angle $\theta$ characterizes the relative
contribution of the scalar complex-power constituent and the CPU-vector
constituent to the CVP norm. It is defined by}
\begin{equation}
\cos\theta
=
\frac{|P+jQ|}{\|\mathbf{S}\|},
\qquad
\sin\theta
=
\frac{\|\mathbf{D}\|}{\|\mathbf{S}\|},
\qquad
0\leq\theta\leq\frac{\pi}{2}.
\label{eq:theta_angle}
\end{equation}
\rev{Equivalently, for $|P+jQ|\neq0$,}
\begin{equation}
\tan\theta
=
\frac{\|\mathbf{D}\|}{|P+jQ|}.
\label{eq:theta_tangent}
\end{equation}
\rev{Therefore, $\theta=0$ when $\mathbf{D}=\mathbf{0}$, whereas $\theta=\pi/2$ when the scalar complex-power constituent vanishes and the CVP norm is entirely associated with the cross-phase constituent, a limiting case illustrated later in Example~1.}

\rev{These angular definitions also provide a geometric interpretation of the compensation criteria developed in Section~\ref{subsec:compensation_criterion}. Conventional power-factor correction adjusts the scalar quadrature component $Q$ and therefore modifies $\varphi$. By contrast, reducing the CPU-vector contribution decreases $\theta$ and makes the equivalent voltage and current vectors more nearly complex-proportional. The CVP therefore distinguishes two complementary compensation directions: modification of the scalar $P$--$Q$ relation and modification of the cross-phase voltage--current structure.
}

\rev{For the configurations considered in this paper, the structured representation is consistent with the additive behavior of the dot- and cross-product terms at a common three-phase terminal.} In particular, for parallel loads sharing the same phase-voltage vector, both the dot-product term $P+jQ$ and the cross-phase term $\mathbf{D}$ add linearly with the corresponding current contributions. \rev{The possible extension of this property to more general network configurations is beyond the scope of the present paper.}

\subsection{Norm Preservation of CVP under Symmetrical Components}

Since the CVP in~\eqref{eq:CVP} is defined in the phase domain, it is necessary to establish whether its properties are preserved when voltages and currents are expressed in a different coordinate system. For the analysis of unbalanced three-phase systems, the natural choice is the symmetrical-components (SC) transformation introduced by Fortescue.

Using the power-invariant form, the phase-domain phasors are mapped as
\begin{equation}
\mathbf{V}^{\pm0}=\mathbf{A}\mathbf{V},
\qquad
\mathbf{I}^{\pm0}=\mathbf{A}\mathbf{I},
\end{equation}
with
\begin{equation}
\mathbf{A}=\frac{1}{\sqrt{3}}
\begin{bmatrix}
1 & a & a^{2}\\
1 & a^{2} & a\\
1 & 1 & 1
\end{bmatrix},
\qquad
a=e^{j2\pi/3}.
\end{equation}

This matrix is unitary. Under the notation adopted here, the corresponding
power-invariance condition is
\begin{equation}
\mathbf{A}^{T}\mathbf{A}^{*}=\mathbf{I},
\end{equation}
which directly implies preservation of the dot product,
\begin{equation}
\mathbf{V}^{\pm0}\cdot\left(\mathbf{I}^{\pm0}\right)^{*}
=\mathbf{V}\cdot\mathbf{I}^{*}.
\end{equation}

Consequently, the classical complex power $P+jQ$---including contributions from positive-, negative-, and zero-sequence components---remains unchanged under the symmetrical-components transformation.

The cross product requires special attention, since it is not preserved under a general linear mapping. However, for the Fortescue matrix it can be shown that
\begin{equation}
(\mathbf{A}\mathbf{V})\times(\mathbf{A}\mathbf{I})
=\det(\mathbf{A})\,\mathbf{A}^{*}(\mathbf{V}\times\mathbf{I}).
\end{equation}

\rev{The derivation is given in Appendix~B.}

Because $|\det(\mathbf{A})|=1$ and $\mathbf{A}$ is unitary, the transformation preserves the norm of the cross component. As a result,
\begin{equation}
\mathbf{D}^{\pm0}=\det(\mathbf{A})\,\mathbf{A}^{*}\mathbf{D},
\qquad
\|\mathbf{D}^{\pm0}\|=\|\mathbf{D}\|.
\end{equation}

Since both the dot and cross components maintain their norms, the norm of the CVP is preserved,
\begin{equation}
\|\mathbf{S}\|^{2}
=
\left|
\mathbf{V}^{\pm0}\cdot
\left(\mathbf{I}^{\pm0}\right)^{*}
\right|^{2}
+\|\mathbf{V}^{\pm0}\times\mathbf{I}^{\pm0}\|^{2}
=
\left|\mathbf{V}\cdot\mathbf{I}^{*}\right|^{2}
+\|\mathbf{V}\times\mathbf{I}\|^{2}.
\end{equation}

\rev{Thus, the symmetrical-components transformation changes the coordinate representation while preserving the CVP metric structure.}

In three-phase three-wire (3P--3W) systems, Kirchhoff's current law forces the zero-sequence current to vanish ($\bar I^{0}=0$), yielding
\begin{equation}
P=P^{+}+P^{-}, \qquad
Q=Q^{+}+Q^{-}.
\end{equation}

However, a nonzero zero-sequence voltage may still exist; in such cases, although $\bar I^{0}=0$, the cross component $\mathbf{D}^{\pm0}$ generally retains a three-dimensional structure. If, in addition, $\bar V^{0}=0$, both voltage and current vectors are confined to the $(+,-)$ plane, even when positive- and negative-sequence components coexist, and their cross product is restricted to a single direction normal to this plane.

This possibility of dimensional reduction motivates the use of the symmetrical-components basis as an intermediate analytical space for extending the formulation to three-phase four-wire systems, addressed in the next section.

\section{Extension of the 3P--3W CVP to 3P--4W Systems}

Three-phase four-wire (3P--4W) networks require an explicit and properly defined voltage reference, since the presence of a neutral conductor introduces an additional degree of freedom that is not naturally accommodated by the three-coordinate voltage and current representation underlying the CVP formulation. While Kirchhoff's current law limits the number of independent line currents to three, no analogous constraint exists for voltages in a four-terminal configuration; the reference must therefore be specified so as to satisfy a suitable compatibility condition.

Measurements at the point of observation (POC) provide the phase voltages
\begin{equation}
\mathbf{V}=
\begin{bmatrix}
\bar V_{1N}\\[2pt]
\bar V_{2N}\\[2pt]
\bar V_{3N}
\end{bmatrix},
\end{equation}
but these quantities depend on the adopted reference. To obtain a three-coordinate representation consistent with the CVP framework and with the effective-apparent-power formulation of IEEE~1459, the phase voltages are referred to an artificial (virtual) neutral $O$ defined by the generalized barycentric condition
\begin{equation}
\bar V_{1O}+\bar V_{2O}+\bar V_{3O}+\frac{\bar V_{NO}}{\rho}=0,
\end{equation}
with neutral-to-phase resistance ratio $\rho=R_N/R_S$. This condition specifies the uniform shift that maps the measured line-to-neutral voltages onto the line-to-$O$ voltages,
\begin{equation}
\mathbf{V}_{O}=\mathbf{V}+\bar V_{NO}\,\mathbf{1},
\qquad
\mathbf{1}=
\begin{bmatrix}
1\\[2pt]1\\[2pt]1
\end{bmatrix},
\end{equation}
where the shift $\bar V_{NO}$ follows explicitly as
\begin{equation}
\bar V_{NO}
=
-\frac{(\bar V_{1N}+\bar V_{2N}+\bar V_{3N})}{\,3+\frac{1}{\rho}\,}.
\end{equation}
This construction restores a three-coordinate voltage representation compatible with the CVP framework.

The shifted voltages and line currents are then expressed in symmetrical components using the power-invariant Fortescue matrix $\mathbf{A}$,
\begin{equation}
\mathbf{V}^{\pm0}_{O}=\mathbf{A}\mathbf{V}_{O},
\qquad
\mathbf{I}^{\pm0}=\mathbf{A}\mathbf{I}.
\end{equation}

To ensure consistency with IEEE Std.~1459 under sinusoidal steady-state conditions, a metric correction is applied to the zero-sequence components within the CVP framework, defining the equivalent quantities
\begin{equation}
\bar V^{0}_{e}=\sqrt{1+3\rho}\,\bar {V}^{0}_{O},
\qquad
\bar I^{0}_{e}=\sqrt{1+3\rho}\,\bar{I}^{0}.
\end{equation}
The resulting equivalent sequence vectors are
\begin{equation}
\mathbf{V}^{\pm0}_{e}=
\begin{bmatrix}
\bar {V}^{+}_{O} & \bar V^{-}_{O} & \bar V^{0}_{e}
\end{bmatrix}^{\mathsf{T}},
\qquad
\mathbf{I}^{\pm0}_{e}=
\begin{bmatrix}
\bar I^{+} & \bar I^{-} & \bar I^{0}_{e}
\end{bmatrix}^{\mathsf{T}}.
\end{equation}

In these coordinates, the dot and cross components of the CVP retain their forms,
\begin{equation}
\mathbf{V}^{\pm0}_{e}\cdot
\left(\mathbf{I}^{\pm0}_{e}\right)^{*}
=P+jQ,
\qquad
\mathbf{V}^{\pm0}_{e}\times\mathbf{I}^{\pm0}_{e}=\mathbf{D}^{\pm0}_{e},
\end{equation}
and the norm of the cross-phase unbalance component is preserved within the equivalent metric.

Applying the inverse transformation provides equivalent phase-domain phasors
\begin{equation}
\mathbf{V}_{e}=\mathbf{A}^{-1}\mathbf{V}^{\pm0}_{e},
\qquad
\mathbf{I}_{e}=\mathbf{A}^{-1}\mathbf{I}^{\pm0}_{e},
\end{equation}
which, in terms of the measured line-to-neutral phasors, admit the explicit expressions
\begin{equation}
\mathbf{V}_{e}
=
\mathbf{V}_{O}
-
k(\rho)\,\bar V_{NO}\,\mathbf{1},
\qquad
\mathbf{I}_{e}
=
\mathbf{I}
+
\rho\,k(\rho)\,\bar I_{N}\,\mathbf{1},
\label{eq:Ve_Ie_explicit}
\end{equation}
where
\begin{equation}
\bar I_{N}=\bar I_{1}+\bar I_{2}+\bar I_{3},
\qquad
k(\rho)=\frac{\sqrt{\,1+3\rho\,}-1}{3\rho}.
\end{equation}

To avoid ambiguity, we note that the subscript $e$ is used in two complementary ways: when attached to scalars ($V_e$, $I_e$, $S_e$), it denotes the effective rms quantities defined in IEEE~1459; when attached to the vectors $\mathbf{V}_e$ and $\mathbf{I}_e$, it denotes the equivalent CVP coordinates whose norms reproduce the corresponding effective rms values. The notation $\mathbf{S}_e$ denotes the resulting structured CVP representation.

These equivalents preserve the CVP metric and power structure:
\begin{equation}
\begin{aligned}
\|\mathbf{V}_{e}\| &= \|\mathbf{V}^{\pm0}_{e}\|, \quad
\|\mathbf{I}_{e}\| = \|\mathbf{I}^{\pm0}_{e}\|
\end{aligned}
\label{eq:metric-preserve-VI}
\end{equation}
\begin{equation}
\begin{aligned}
\mathbf{V}_{e}\cdot\mathbf{I}_{e}^{*} &= P + jQ, \quad
\|\mathbf{D}_{e}\| = \|\mathbf{V}_{e}\times\mathbf{I}_{e}\| = \|\mathbf{D}^{\pm0}_{e}\|
\end{aligned}
\label{eq:metric-preserve-power}
\end{equation}

Consequently,
\begin{equation}
\|\mathbf{S}_{e}\|
=
\sqrt{P^{2}+Q^{2}+\|\mathbf{D}_{e}\|^{2}}
=
\|\mathbf{S}^{\pm0}_{e}\|.
\end{equation}

Moreover, given that the equivalent voltage and current vectors have been constructed to match the corresponding effective rms quantities prescribed by IEEE~1459 in this regime, the equivalent apparent-power norm likewise coincides, by construction, with the IEEE~1459 effective apparent power~$S_e$.

\subsection*{Limiting operating conditions}

\subsubsection*{Balanced supply}
If the supply is balanced, the artificial-neutral shift vanishes ($\bar V_{NO}=0$) and the zero-sequence voltage is zero ($\bar V^{0}_{O}=0$). In 3P--4W systems, however, unbalanced loads may still give rise to a nonzero zero-sequence current ($\bar I^{0}\neq 0$) through the neutral. Since $\bar V^{0}_{O}=0$, this zero-sequence current component does not affect the dot-product term $P+jQ$, but still plays a role in the cross-phase structure captured by $\mathbf{D}_{e}$.

\subsubsection*{Identical conductor resistances ($\rho=1$)}
The generalized barycentric condition reduces to the classical virtual-star-point condition of the FBD method~\cite{Depenbrock1993}:
\begin{equation}
\bar V_{1O}+\bar V_{2O}+\bar V_{3O}+\bar V_{NO}=0,
\end{equation}
which implies the uniform shift
\begin{equation}
\bar V_{NO}= -\frac{(\bar V_{1N}+\bar V_{2N}+\bar V_{3N})}{4}.
\end{equation}
In the equivalent-phase construction one has $k(1)=\dfrac{1}{3}$.

\subsubsection*{Perfect neutral ($\rho=0$)}
The physical and artificial neutrals coincide ($N\equiv O$), hence $\bar V_{NO}=0$ and
$\mathbf{V}_{e}=\mathbf{V}$, $\mathbf{I}_{e}=\mathbf{I}$.
Although $k(\rho)\to \tfrac{1}{2}$, the correction is immaterial because it multiplies a vanishing shift.

\subsubsection*{Three-phase three-wire ($\rho\to\infty$)}
The artificial-neutral shift tends to the barycentric average
\begin{equation}
\bar V_{NO}= -\frac{(\bar V_{1N}+\bar V_{2N}+\bar V_{3N})}{3},
\end{equation}
the zero-sequence current vanishes ($\bar I^{0}=0$), and $k(\rho)\to 0$.
The formulation collapses to the standard 3P--3W model based on positive- and negative-sequence components.

\section{Instantaneous Representation of the Cross-Phase Term}

Consider a three-phase system operating under sinusoidal steady-state conditions, with instantaneous phase voltages and currents expressed with respect to the adopted POC reference. Although different reference choices would modify the numerical values of the phase quantities and the resulting cross-product terms, the algebraic structure of the CVP relations remains unchanged. The instantaneous voltages and currents are therefore written as
\begin{equation}
v_k(t)=\sqrt{2}\,V_k\sin(\omega t+\alpha_k), \qquad
i_k(t)=\sqrt{2}\,I_k\sin(\omega t+\beta_k),
\end{equation}

These instantaneous quantities define not only the scalar instantaneous power
\begin{equation}
p(t)=\sum_{k=1}^{3} v_k(t)i_k(t),
\end{equation}
but also the antisymmetric term
\begin{equation}
\mathbf{d}(t)\triangleq \mathbf{v}(t)\times\mathbf{i}(t),
\end{equation}
with
\begin{equation}
\mathbf{v}(t)=
\begin{bmatrix}
v_1(t)\\ v_2(t)\\ v_3(t)
\end{bmatrix},
\qquad
\mathbf{i}(t)=
\begin{bmatrix}
i_1(t)\\ i_2(t)\\ i_3(t)
\end{bmatrix}.
\end{equation}

Its components are the antisymmetric combinations
\begin{align}
d_1(t)&=v_2(t)i_3(t)-v_3(t)i_2(t), \nonumber\\
d_2(t)&=v_3(t)i_1(t)-v_1(t)i_3(t), \nonumber\\
d_3(t)&=v_1(t)i_2(t)-v_2(t)i_1(t),
\end{align}
an arrangement also considered in~\cite{Peng1996} from a different perspective.

Decomposing $\mathbf{d}(t)$ into mean and oscillatory parts,
\begin{equation}
\tilde{\mathbf{d}}(t)=\mathbf{d}(t)-\langle\mathbf{d}\rangle,
\end{equation}
each component of $\tilde{\mathbf{d}}(t)$ is a double-frequency sinusoid,
\begin{equation}
\tilde d_k(t)=D_k \cos(2\omega t+\psi_{dk}),
\qquad k=1,2,3.
\end{equation}

The amplitudes and phase angles are determined by the cross-product coefficients
\begin{equation}
\bar D_k=\bar V_\ell \bar I_m-\bar V_m \bar I_\ell,
\qquad
D_k=|\bar D_k|,
\qquad
\psi_{dk}=\arg(\bar D_k),
\end{equation}
where $(k,\ell,m)$ denotes any cyclic permutation of $(1,2,3)$. Hence, these coefficients coincide with the components of the phasor cross product
\begin{equation}
\mathbf{D}=\mathbf{V}\times\mathbf{I},
\end{equation}
i.e., the CPU term defined in Section~II.

The vector $\mathbf{D}$ therefore represents the amplitude and phase of the corresponding $2\omega$ oscillatory terms, in parallel with Steinmetz's observation that products of alternating quantities generate double-frequency components~\cite{Steinmetz1899}.

Since each $\tilde d_k(t)$ has zero mean, its amplitude is characterized by its rms value,
\begin{equation}
\sigma_k=
\sqrt{\frac{1}{T}\int_{0}^{T} \tilde d_k^2(t)\,dt}
=
\frac{D_k}{\sqrt{2}}.
\end{equation}

Accordingly, the rms magnitude of the full cross-phase term is
\begin{equation}
\sigma_d=
\sqrt{\sum_{k=1}^{3}\sigma_k^2}
=
\frac{\|\mathbf{D}\|}{\sqrt{2}}.
\end{equation}

Hence, the norm $\|\mathbf{D}\|$ obtained from the phasor cross product is directly related to the rms magnitude of its instantaneous counterpart. The cross-phase term represents the antisymmetric inter-phase correlation between voltages and currents.

\section{Numerical Examples}

\rev{This section first presents two numerical configurations illustrating the application of the proposed CVP to three-phase four-wire systems under sinusoidal steady-state conditions. The second configuration is subsequently used to validate the compensation criteria developed in Section~\ref{subsec:compensation_criterion}. Uncoupled lumped elements are used for clarity, but this choice is not a restriction of the CVP formulation itself. The reported quantities include the phase-domain phasors, the corresponding equivalent coordinates $(\mathbf{V}_e,\mathbf{I}_e)$, the scalar term $P+jQ$, the CPU vector $\mathbf{D}_e$, and the resulting apparent-power norm and power factor.}

%------------------------------------------------

\subsection{Example~1 --- Purely Reactive Asymmetric Load}

Example~1 presents a conceptual configuration intended to illustrate the structural behavior of the proposed formulation under ideal conditions. A balanced three-phase source feeds a star-connected load composed of three purely reactive impedances expressed in per-unit with respect to a common reactance $X$. The load neutral is directly connected to the source neutral. The per-unit system uses $V$ as the voltage base, $I=V/X$ as the current base, and $VI$ as the base for all power components. In particular, the phase impedances are
\begin{equation}
\bar Z_1=jX,\qquad 
\bar Z_2=-j5X,\qquad 
\bar Z_3=-j1.25X.
\end{equation}

The system is fed by balanced line-to-neutral voltage phasors
\begin{equation}
\mathbf{V}=
\begin{bmatrix}
V & V\angle -120^\circ & V\angle 120^\circ
\end{bmatrix}^{\mathsf T}.
\end{equation}

Since the supply is symmetrical, the artificial-neutral shift is zero ($\bar V_{NO}=0$), and the phase voltages coincide with those referred to the virtual neutral.

Although each branch is purely reactive, the load is strongly asymmetric. The resulting phase currents are therefore unbalanced and produce a nonzero neutral current ($\bar I_N \neq 0$). \rev{Accordingly, for $\rho=1$, the equivalent current vector $\mathbf{I}_e$ is obtained from~\eqref{eq:Ve_Ie_explicit}.}

The scalar powers satisfy
\begin{equation}
P=0,\qquad Q=0,
\end{equation}
Although the total reactive power is zero, the individual phase contributions
remain nonzero and cancel algebraically, while the cross-phase unbalance vector
remains nonzero,
\begin{equation}
\mathbf{D}_{e}\neq\mathbf{0}.
\end{equation}

In this configuration, the CVP is entirely determined by the cross-phase term; consequently, the apparent power reduces to
\begin{equation}
\|\mathbf{S}_{e}\|
=
\|\mathbf{D}_{e}\|
=
\|\mathbf{V}_e\|\,\|\mathbf{I}_e\|.
\end{equation}

This example illustrates that the apparent-power norm may remain nonzero even when the dot-product contribution $P+jQ$ vanishes, provided that the phase currents are not symmetrically distributed. Within the CVP framework, this residual apparent power is entirely associated with cross-phase unbalance.

The corresponding phasors and power quantities are reported in Tables~\ref{tab:ex1_phasors} and~\ref{tab:ex1_powers}.

%------------------------------------------------
\begin{table}[!t]
\caption{Example 1: Phasor Vectors and Norms (per-unit values)}
\label{tab:ex1_phasors}
\centering
\begin{tabular}{c c}
\hline
\\[-7pt]
Quantity & Example 1 \\[1pt]
\hline\\[-6pt]
$\mathbf{V}=\mathbf{V}_e$ &
$\begin{bmatrix}
1\\
1\angle -120^\circ\\
1\angle 120^\circ
\end{bmatrix}$ \\[12pt]

$\mathbf{V}^{\pm0}_{e}$ &
$\begin{bmatrix}
\sqrt{3}\\
0\\
0
\end{bmatrix}$ \\[14pt]

$\|\mathbf{V}\|$ & $\sqrt{3}$ \\[6pt]

$\mathbf{I}$ &
$\begin{bmatrix}
1\angle -90^\circ\\
0.2\angle -30^\circ\\
0.8\angle -150^\circ
\end{bmatrix}$ \\[14pt]

$\bar{I}_N$ & $\dfrac{\sqrt{63}}{5}\angle -109.107^\circ$ \\[8pt]

$\mathbf{I}_e$ &
$\begin{bmatrix}
\dfrac{\sqrt{57}}{5}\angle -96.587^\circ\\
0.6\angle -90^\circ\\
\dfrac{\sqrt{39}}{5}\angle -133.898^\circ
\end{bmatrix}$ \\[12pt]

$\mathbf{I}^{\pm0}$ &
$\begin{bmatrix}
0\\
\dfrac{\sqrt{21}}{5}\angle -70.893^\circ\\
\dfrac{\sqrt{21}}{5}\angle -109.107^\circ
\end{bmatrix}$ \\[12pt]

$\mathbf{I}^{\pm0}_{e}$ &
$\begin{bmatrix}
0\\
\dfrac{\sqrt{21}}{5}\angle -70.893^\circ\\
\dfrac{2\sqrt{21}}{5}\angle -109.107^\circ
\end{bmatrix}$ \\[14pt]

$\|\mathbf{I}^{\pm0}_{e}\|=\|\mathbf{I}_e\|$ &
$\dfrac{\sqrt{105}}{5}$ \\[8pt]
\hline
\end{tabular}
\end{table}
%------------------------------------------------

%------------------------------------------------
%------------------------------------------------
\begin{table}[!t]
\caption{Example 1: Power Components and Norms (per-unit values)}
\label{tab:ex1_powers}
\centering
\begin{tabular}{c c}
\hline\\[-7pt]
Power Component & Example 1 \\[1pt]
\hline\\[-6pt]
$P$ & $0$ \\[8pt]

$Q$ & $0$ \\[8pt]

$\mathbf{D}_e$ &
$\begin{bmatrix}
\dfrac{\sqrt{39}}{5}\angle 133.898^\circ\\
\dfrac{\sqrt{183}}{5}\angle 33.67^\circ\\
\dfrac{\sqrt{93}}{5}\angle -51.052^\circ
\end{bmatrix}$ \\[12pt]

$\mathbf{D}^{\pm0}_{e}$ &
$\begin{bmatrix}
0\\
\dfrac{2\sqrt{63}}{5}\angle 70.893^\circ\\
\dfrac{\sqrt{63}}{5}\angle \rev{-70.893^\circ}
\end{bmatrix}$ \\[16pt]

$\|\mathbf{D}_{e}\| = \|\mathbf{S}_{e}\|$ & $\dfrac{3\sqrt{35}}{5}$ \\[8pt]
\hline
\end{tabular}
\end{table}
%------------------------------------------------

\subsection{Example~2 --- Unbalanced supply with unequal load and conductor resistances}

Example~2 considers a more realistic four-wire distribution configuration in which supply asymmetry, load unbalance, and unequal conductor parameters are simultaneously present, inspired by the circuit configuration discussed in \cite[p.~175]{Emanuel2011}. The system is modeled as a three-phase feeder with finite conductor impedances. The phase conductors have $R_S=1.0~\Omega$ and $L_S=7.6$~mH, while the neutral conductor has the same inductance $L_N=L_S$ but a higher resistance $R_N=2.40~\Omega$, yielding $\rho=2.4$.

\begin{figure}[!t]
\centering
\includegraphics[width=\columnwidth]{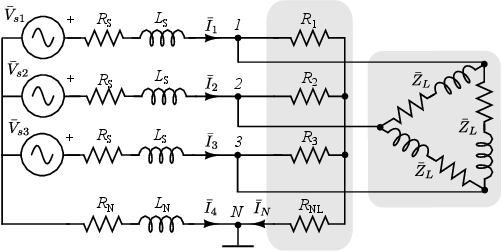}
\caption{Example~2: Unbalanced supply with unequal load and conductor resistances.}
\label{fig:ex2_circuit}
\end{figure}

The load consists of a strongly unbalanced line-to-neutral resistive set
$R_1=100~\Omega$, $R_2=500~\Omega$, and $R_3=1000~\Omega$, in parallel with a balanced delta-connected impedance
$\bar Z_L=70+j70.48~\Omega$. On the load side, the neutral conductor includes $R_{NL}=0.04~\Omega$ (Fig.~\ref{fig:ex2_circuit}).

The source provides asymmetric 60-Hz line-to-neutral voltages. Using the artificial-neutral reference and the proposed equivalent coordinates, the resulting power quantities are consistent with the effective apparent-power definition prescribed in IEEE Std.~1459:
\begin{equation}
P=648.655~\text{W},\qquad
Q=542.717~\text{var},
\end{equation}
\begin{equation}
\|\mathbf{D}_{e}\|=228.403~\text{VA},\qquad
\|\mathbf{S}_{e}\|=876.05~\text{VA},
\qquad
\text{PF}=0.74.
\end{equation}

Computation in the modified symmetrical-components domain yields identical dot-product and cross-product norms, confirming preservation of the CVP metric under the coordinate
transformation. The resulting phasors and power quantities for Example~2 are reported in Tables~\ref{tab:ex2_phasors} and~\ref{tab:ex2_power}.

%------------------------------------------------

\begin{table}[!t]
\caption{Example 2 ($\rho=2.4$): Phase and Equivalent Phasors (V, A).}
\label{tab:ex2_phasors}
\centering
\setlength{\tabcolsep}{3pt}
\renewcommand{\arraystretch}{0.95}
\begin{tabular}{c c}
\hline\\[-7pt]
Quantity & Example 2 \\[1pt]
\hline\\[-6pt]

$\mathbf{V}$ &
$\begin{bmatrix}
91.50\angle -5.50^\circ\\
94.78\angle -123.81^\circ\\
89.62\angle 121.25^\circ
\end{bmatrix}$ \\[12pt]

$\bar{V}_{NO}$ &
$3.985\angle 53.214^\circ$ \\[6pt]

$\mathbf{V}_O$ &
$\begin{bmatrix}
93.63\angle -3.42^\circ\\
90.80\angle -123.68^\circ\\
91.19\angle 118.93^\circ
\end{bmatrix}$ \\[12pt]

$k(\rho)$ & $\rev{0.259}$ \\[6pt]

$\mathbf{V}_e$ &
$\begin{bmatrix}
93.07\angle -3.95^\circ\\
91.83\angle -123.71^\circ\\
\rev{90.77\angle 119.52^\circ}
\end{bmatrix}$ \\[12pt]

$\mathbf{V}^{\pm0}_{e}$ &
$\begin{bmatrix}
159.10\angle -2.73^\circ\\
3.79\angle -20.47^\circ\\
2.75\angle \rev{-126.79^\circ}
\end{bmatrix}$ \\[12pt]

$\|\mathbf{V}^{\pm0}_{e}\|=\|\mathbf{V}_e\|$ & $159.163$ \\[8pt]

$\mathbf{I}$ &
$\begin{bmatrix}
3.562\angle -38.28^\circ\\
2.863\angle -166.17^\circ\\
2.822\angle 74.76^\circ
\end{bmatrix}$ \\[12pt]

$\bar{I}_N$ &
$0.776\angle -12.52^\circ$ \\[6pt]

$\mathbf{I}_e$ &
$\begin{bmatrix}
4.0\angle -35.28^\circ\\
2.44\angle -161.14^\circ\\
2.89\angle 65.15^\circ
\end{bmatrix}$ \\[12pt]

$\mathbf{I}^{\pm0}_{e}$ &
$\begin{bmatrix}
5.33\angle -42.85^\circ\\
0.51\angle -11.5^\circ\\
1.28\angle \rev{-12.52^\circ}
\end{bmatrix}$ \\[12pt]

$\|\mathbf{I}^{\pm0}_{e}\|=\|\mathbf{I}_e\|$ & $5.504$ \\[2pt]
\hline
\end{tabular}
\end{table}
%------------------------------------------------

%------------------------------------------------
%------------------------------------------------
\begin{table}[!t]
\caption{Example 2 ($\rho=2.4$): Power Components and Norms (W, var, VA).}
\label{tab:ex2_power}
\centering
\begin{tabular}{c c}
\hline\\[-7pt]
Power Component & Example 2 \\[1pt]
\hline\\[-6pt]

$P$ & $648.66$ \\[6pt]

$Q$ & $542.72$ \\[6pt]

$\mathbf{D}_e$ &
$\begin{bmatrix}
83.6\angle -109.13^\circ\\
156.62\angle 126.39^\circ\\
143.7\angle 30.66^\circ
\end{bmatrix}$ \\[12pt]

$\mathbf{D}^{\pm0}_{e}$ &
$\begin{bmatrix}
5.4\angle -18.52^\circ\\
217.5\angle 166.42^\circ\\
69.53\angle -1.57^\circ
\end{bmatrix}$ \\[12pt]

$\|\mathbf{D}_{e}\|=\|\mathbf{D}^{\pm0}_{e}\|$ & $228.40$ \\[6pt]

$\|\mathbf{S}_{e}\|$ & $876.05$ \\[6pt]

PF & $0.74$ \\
\hline
\end{tabular}
\end{table}
%------------------------------------------------
%------------------------------------------------
\rev{
\subsection{CVP-Based Compensation Criterion}
\label{subsec:compensation_criterion}

The following criterion is introduced only to assess the practical use of the CPU descriptor. It is not intended as a complete converter-control strategy or as an optimization procedure. Its purpose is to show that the CVP quantities evaluated from terminal phasors can be translated into reproducible shunt-current references.

Let $\mathbf{V}_e$ denote the equivalent rms voltage vector at the observation point. For prescribed active- and reactive-power values $P_{\mathrm{ref}}$ and $Q_{\mathrm{ref}}$, the equivalent current reference
\begin{equation}
\mathbf{I}_{e,\mathrm{ref}}
=
\frac{P_{\mathrm{ref}}-jQ_{\mathrm{ref}}}
{\|\mathbf{V}_e\|^2}\,
\mathbf{V}_e
\label{eq:Ie_ref_collinear}
\end{equation}
satisfies
\begin{equation}
\mathbf{V}_e\cdot
\mathbf{I}_{e,\mathrm{ref}}^{*}
=
P_{\mathrm{ref}}+jQ_{\mathrm{ref}},
\qquad
\mathbf{V}_e\times
\mathbf{I}_{e,\mathrm{ref}}
=
\mathbf{0}.
\label{eq:Ie_ref_collinear_properties}
\end{equation}
It therefore retains the prescribed scalar complex-power term while imposing the limiting CPU-cancellation condition. This condition makes the equivalent voltage and current vectors complex-proportional, without implying that either phasor set is individually balanced.

For a 3P--4W system, define
\begin{equation}
c(\rho)=\rho k(\rho).
\label{eq:c_rho_comp}
\end{equation}
Inversion of the equivalent-current relation in~\eqref{eq:Ve_Ie_explicit} gives the corresponding physical phase-current reference:
\begin{equation}
\mathbf{I}_{\mathrm{ref}}
=
\mathbf{I}_{e,\mathrm{ref}}
-
\frac{c(\rho)}
{1+3c(\rho)}
\left(
\mathbf{1}^{T}\mathbf{I}_{e,\mathrm{ref}}
\right)\mathbf{1}.
\label{eq:Ie_inverse_comp}
\end{equation}
If $\mathbf{I}_{\mathrm{L}}$ denotes the aggregate physical load-current vector and positive compensator current is defined as injected from the neutral toward the phases, the required shunt-current reference is
\begin{equation}
\mathbf{I}_{\mathrm{c,ref}}
=
\mathbf{I}_{\mathrm{L}}
-
\mathbf{I}_{\mathrm{ref}},
\label{eq:Ic_ref}
\end{equation}
and the source-side current is
\begin{equation}
\mathbf{I}_{\mathrm{S}}
=
\mathbf{I}_{\mathrm{L}}
-
\mathbf{I}_{\mathrm{c}}.
\label{eq:Is_load_comp}
\end{equation}

A second use of the CVP follows directly from
\begin{equation}
\mathrm{PF}
=
\frac{P}
{\sqrt{P^2+Q^2+\|\mathbf{D}_e\|^2}}.
\label{eq:PF_comp}
\end{equation}
For a prescribed power factor $\mathrm{PF}_{\mathrm{ref}}$ and active-power reference $P_{\mathrm{ref}}$, the corresponding nonactive margin is
\begin{equation}
M_{\mathrm{ref}}
=
|P_{\mathrm{ref}}|
\sqrt{
\frac{1}{\mathrm{PF}_{\mathrm{ref}}^2}-1
}.
\label{eq:M_ref}
\end{equation}
An allocation parameter $\eta\in[0,1]$ may then be used to prescribe
\begin{equation}
Q_{\mathrm{ref}}
=
s_Q\sqrt{\eta}\,M_{\mathrm{ref}},
\qquad
D_{\mathrm{ref}}
=
\sqrt{1-\eta}\,M_{\mathrm{ref}},
\label{eq:eta_allocation}
\end{equation}
where $D_{\mathrm{ref}}$ is the target value of
$\|\mathbf{D}_e\|$ and $s_Q\in\{-1,1\}$ specifies the selected sign of
reactive power. The parameter $\eta$ is an allocation parameter and not an optimization weight.

To construct the corresponding reference, the measured equivalent load-current vector is decomposed as
\begin{equation}
\mathbf{I}_{e,\mathrm{L}}
=
\frac{P_{\mathrm{L}}-jQ_{\mathrm{L}}}
{\|\mathbf{V}_e\|^2}\,
\mathbf{V}_e
+
\mathbf{I}_{e,\mathrm{L}}^{\perp},
\label{eq:Ie_load_decomposition}
\end{equation}
where
\begin{equation}
\mathbf{V}_e\cdot
\left(
\mathbf{I}_{e,\mathrm{L}}^{\perp}
\right)^{*}
=
0.
\label{eq:Ie_perp_property}
\end{equation}
The equivalent current reference is then
\begin{equation}
\begin{aligned}
\mathbf{I}_{e,\mathrm{ref}}
={}&
\frac{P_{\mathrm{ref}}-jQ_{\mathrm{ref}}}
{\|\mathbf{V}_e\|^2}\,
\mathbf{V}_e
+
\gamma\mathbf{I}_{e,\mathrm{L}}^{\perp},
\\
\gamma
={}&
\frac{D_{\mathrm{ref}}}
{\|\mathbf{D}_{e,\mathrm{L}}\|},
\qquad
0\leq\gamma\leq1.
\end{aligned}
\label{eq:Ie_ref_allocation}
\end{equation}
This construction yields
\begin{equation}
\mathbf{V}_e\cdot
\mathbf{I}_{e,\mathrm{ref}}^{*}
=
P_{\mathrm{ref}}+jQ_{\mathrm{ref}},
\qquad
\mathbf{D}_{e,\mathrm{ref}}
=
\gamma\mathbf{D}_{e,\mathrm{L}}.
\label{eq:Ie_ref_allocation_properties}
\end{equation}
Consequently,
\begin{equation}
\|\mathbf{D}_{e,\mathrm{ref}}\|
=
\gamma\|\mathbf{D}_{e,\mathrm{L}}\|
=
D_{\mathrm{ref}}.
\label{eq:De_ref_norm}
\end{equation}
Thus, the measured CPU vector is scaled without being interpreted as a set of independent physical-power commands. The present construction modifies its norm while retaining its direction in the equivalent-coordinate space.

When the feeder impedances are nonzero, the injected current modifies the terminal voltage and the load operating point. In the numerical assessment below, the reference is therefore recalculated from the updated terminal phasors until the resulting CVP quantities remain within the stated numerical tolerance. This network-dependent update does not alter the algebraic definition of the CVP.

\subsection{Compensation Assessment Using Example~2}
\label{subsec:compensation_assessment}

The circuit of Example~2 is now used to assess whether the CVP quantities can define actionable compensation objectives. The purpose is not to develop the inner control loops or the power-electronic implementation of a particular device, but to verify that terminal phasors can be converted into reproducible shunt-current references that modify $Q$ and $\|\mathbf{D}_e\|$ according to prescribed objectives.

The circuit was implemented in sinusoidal phasor mode at 60~Hz. Three independently controlled single-phase current sources were connected in shunt between the phase conductors and the physical neutral at the load point of common coupling. Positive source current was defined from the neutral toward the corresponding phase. Each ideal current source was paralleled by a numerical regularization resistance of
$R_p=1~\mathrm{M}\Omega$. The resulting resistance currents were negligible with respect to the compensating currents.

The source-side performance was evaluated from the three-phase voltages and currents measured immediately upstream of the parallel combination formed by the loads and the compensator. The corresponding rms phasors were used
to construct the equivalent vectors $\mathbf{V}_e$ and $\mathbf{I}_e$
through~\eqref{eq:Ve_Ie_explicit}. The currents of the star-connected and delta-connected loads were measured separately. With the adopted current directions, the phase-current balance was numerically verified as
\begin{equation}
\mathbf{I}_{\mathrm{S}}
=
\mathbf{I}_{\star}
+
\mathbf{I}_{\Delta}
-
\mathbf{I}_{\mathrm{c}}.
\label{eq:sim_current_balance}
\end{equation}
The current-balance residual,
$\varepsilon_I=
\|\mathbf{I}_{\mathrm{S}}
-\mathbf{I}_{\star}
-\mathbf{I}_{\Delta}
+\mathbf{I}_{\mathrm{c}}\|$,
remained below $1.68\times10^{-4}$~A in the reported compensated operating points. This small residual is consistent with the currents through the numerical regularization resistances.

The initial operating point was that reported in
Tables~\ref{tab:ex2_phasors} and~\ref{tab:ex2_power}:
\begin{equation}
P_0=648.655~\mathrm{W},
\qquad
Q_0=542.717~\mathrm{var},
\label{eq:ex2_initial_PQ}
\end{equation}
\begin{equation}
\|\mathbf{D}_{e,0}\|
=
228.403~\mathrm{VA},
\qquad
\mathrm{PF}_0=0.74.
\label{eq:ex2_initial_DPF}
\end{equation}

\subsubsection{CPU cancellation at prescribed $P$ and $Q$}

The first compensation test used~\eqref{eq:Ie_ref_collinear} with
\begin{equation}
P_{\mathrm{ref}}=P_0,
\qquad
Q_{\mathrm{ref}}=Q_0.
\label{eq:case1_targets}
\end{equation}
Thus, the compensator was required to reduce the CPU contribution while retaining the initial scalar complex-power term. Because the feeder impedances cause the terminal voltage and the load operating point to vary with the injected current, the reference was successively recalculated from the updated terminal phasors until the resulting CVP quantities remained within the adopted numerical tolerance.

The final source-side quantities were
\begin{equation}
P=647.451~\mathrm{W},
\qquad
Q=541.712~\mathrm{var},
\label{eq:case1_final_PQ}
\end{equation}
\begin{equation}
\|\mathbf{D}_e\|
=
0.224~\mathrm{VA},
\qquad
\|\mathbf{S}_e\|
=
844.183~\mathrm{VA},
\label{eq:case1_final_DS}
\end{equation}
and
\begin{equation}
\mathrm{PF}=0.76696.
\label{eq:case1_final_PF}
\end{equation}
The deviations in $P$ and $Q$ from their prescribed values were approximately
$0.19\%$, whereas the CPU norm was reduced by approximately
$99.90\%$. The resulting upstream rms neutral current was
$\bar I_{N,\mathrm{S}}
=-0.00963-j0.01089$~A.

This result verifies that the condition
$\mathbf{I}_{e,\mathrm{ref}}\propto\mathbf{V}_e$
can be translated into physical phase-current commands and used to suppress the source-side CPU contribution without cancelling the prescribed reactive power. It therefore distinguishes CPU compensation from conventional reactive-power correction.

\subsubsection{Joint $Q$--CPU allocation at a prescribed power factor}

The second test used
\begin{equation}
\mathrm{PF}_{\mathrm{ref}}=0.98,
\qquad
\eta=0.5,
\label{eq:case2_targets}
\end{equation}
which specifies an equal quadratic allocation,
\begin{equation}
Q_{\mathrm{ref}}^2
=
D_{\mathrm{ref}}^2.
\label{eq:case2_equal_allocation}
\end{equation}
At each reference update, $P_{\mathrm{ref}}$ was set equal to the active power of the updated load operating point. The resulting source-side quantities are reported in Table~\ref{tab:compensation_results}. The converged solution gave
$Q=103.505$~var and
$\|\mathbf{D}_e\|=102.996$~VA, yielding
$\mathrm{PF}=0.979731$.
The realized allocation parameter was
$\eta_{\mathrm{real}}=0.50247$, and the difference
$|Q-\|\mathbf{D}_e\||$ was only $0.510$~VA. Hence, the prescribed equal quadratic allocation was closely reproduced.

The required compensating-current vector had an rms norm of
$2.986$~A, with a maximum phase-current magnitude of
$1.782$~A. The active power injected by the ideal current sources was
$0.828$~W, corresponding to approximately $0.12\%$ of the source-side active power. Therefore, the prescribed allocation was obtained with negligible active-power exchange in this operating point. The rms current-balance residual was
$1.67\times10^{-4}$~A.

\begin{table}[!t]
\caption{Source-Side Quantities Before and After Compensation}
\label{tab:compensation_results}
\centering
\footnotesize
\setlength{\tabcolsep}{3.2pt}
\renewcommand{\arraystretch}{1.12}
\begin{tabular}{lccccc}
\hline
\shortstack[l]{Operating\\condition}
& \shortstack{$P$\\(W)}
& \shortstack{$Q$\\(var)}
& \shortstack{$\|\mathbf{D}_e\|$\\(VA)}
& \shortstack{$\|\mathbf{S}_e\|$\\(VA)}
& $\mathrm{PF}$ \\
\hline
Uncompensated
& 648.66
& 542.72
& 228.40
& 876.05
& 0.74043 \\
CPU cancellation
& 647.45
& 541.71
& 0.224
& 844.18
& 0.76696 \\
PF allocation, $\eta=0.5$
& 714.16
& 103.51
& 103.00
& 728.93
& 0.97973 \\
\hline
\end{tabular}
\end{table}

For completeness, the same construction was evaluated for several values of the allocation parameter while retaining
$\mathrm{PF}_{\mathrm{ref}}=0.98$.
The current reference was recalculated from the updated terminal phasors after each simulation. The prescribed numerical tolerances were reached after no more than three reference updates in the evaluated cases. This limited parameter sweep is included only to verify the prescribed allocation and is not formulated as an optimization or controller-design study.

Table~\ref{tab:eta_allocation} confirms that increasing $\eta$ progressively transfers the prescribed nonactive margin from the CPU norm to reactive power. The realized allocation parameter closely follows its prescribed value, while the resulting power factor remains close to $0.98$.

\begin{table}[!t]
\caption{Effect of the Allocation Parameter for
$\mathrm{PF}_{\mathrm{ref}}=0.98$}
\label{tab:eta_allocation}
\centering
\footnotesize
\setlength{\tabcolsep}{2.3pt}
\renewcommand{\arraystretch}{1.10}
\begin{tabular}{ccccccc}
\hline
$\eta$
& $\eta_{\mathrm{real}}$
& $Q$ (var)
& $\|\mathbf{D}_e\|$ (VA)
& $\mathrm{PF}$
& $\|\mathbf{I}_{\mathrm{c}}\|$ (A)
& $P_{\mathrm{c,inj}}$ (W) \\
\hline
$0.2$
& $0.20230$
& $65.84$
& $130.74$
& $0.97993$
& $3.216$
& $0.194$
\\
$0.5$
& $0.50247$
& $103.51$
& $103.00$
& $0.97973$
& $2.986$
& $0.828$
\\
$0.8$
& $0.79544$
& $128.42$
& $65.12$
& $0.98017$
& $2.844$
& $-0.104$
\\
$1.0$
& $0.99962$
& $143.17$
& $2.79$
& $0.98032$
& $2.785$
& $0.090$
\\
\hline
\end{tabular}
\end{table}

The limiting case $\eta=1$ prescribes
$D_{\mathrm{ref}}=0$. The small residual value
$\|\mathbf{D}_e\|=2.79$~VA obtained in the network simulation corresponds to
$\eta_{\mathrm{real}}=0.99962$ and reflects the finite numerical stopping tolerance and the network-dependent change in the terminal operating point.

Because the ideal controlled current sources are not constrained to zero active-power exchange, the compensator active power
\begin{equation}
P_{\mathrm{c,inj}}
=
\Re\left\{
\mathbf{V}^{T}\mathbf{I}_{\mathrm{c}}^{*}
\right\},
\label{eq:Pc_compensator}
\end{equation}
was also evaluated. Here, $\mathbf{V}$ contains the physical phase-to-neutral rms voltages at the compensator terminals, consistently with the adopted current-injection convention. Positive values represent active-power injection into the network, whereas negative values represent active-power absorption. Over the evaluated allocation cases,
$|P_{\mathrm{c,inj}}|$ did not exceed $0.828$~W, or approximately $0.12\%$ of the corresponding source-side active power. This quantity is reported only to characterize the ideal current reference and to distinguish the present proof of concept from a passive compensator or a converter explicitly constrained to zero active-power exchange.

Unlike compensation approaches based primarily on scalar nonactive-power
resolutions or on the enforcement of balanced positive-sequence source
currents, the present CVP construction does not prescribe a unique
compensated operating condition. FBD-type approaches typically seek an
equivalent resistive load and improved conductor utilization, whereas
IEEE~1459-based compensation may target balanced positive-sequence
currents. These objectives remain relevant and are not superseded here.
The CVP provides an additional degree of selectivity: the CPU contribution
may be reduced while retaining a prescribed reactive power, or the
nonactive margin associated with a prescribed power factor may be
continuously allocated between $Q$ and $\|\mathbf{D}_e\|$. Thus, its main
distinctive feature is not a claim of universal optimality, but the ability
to formulate compensation objectives directly in terms of the mutual
structural relation between the equivalent voltage and current vectors.

The two compensation tests serve complementary purposes. The first
isolates the operational meaning of $\|\mathbf{D}_e\|$ by reducing it
while retaining the prescribed $P+jQ$. The second exploits the complete
CVP norm to coordinate $Q$ and $\|\mathbf{D}_e\|$ under a common
power-factor constraint. Together, they provide a numerical proof of
concept that the CPU vector is not merely an algebraic by-product, but a
reproducible structural quantity that can be incorporated into the
construction of compensating-current references.
}

\section{Discussion}

\rev{The proposed CVP provides a structured interpretation of three-phase apparent
power through the conventional complex power $P+jQ$ and the cross-product
term $\mathbf{D}$, which captures interphase antisymmetry.} These constituents satisfy
\[
\|\mathbf{S}\|^2=P^2+Q^2+\|\mathbf{D}\|^2,
\]
so that the apparent-power magnitude is decomposed into intraphase and cross-phase
contributions.

From an engineering viewpoint, the CVP clarifies sources of power-factor
degradation in unbalanced systems. It can distinguish operating conditions with
similar values of $P$ and $Q$ but different internal phase structure, since part of the apparent-power norm may be associated with the
antisymmetric cross-phase relation through $\|\mathbf{D}\|$. In such cases, reactive-power compensation alone does not
necessarily address the full apparent-power demand. The proposed representation
therefore assists the interpretation of unbalanced feeders, loads, and
compensation strategies by separating intraphase phase-shift effects from
cross-phase unbalance effects.

\rev{The compensation assessment provides a limited proof of concept for
the engineering use of the CPU descriptor. It shows that the CPU contribution
can be incorporated into terminal current references, either through its
selective reduction while retaining prescribed $P$ and $Q$, or through a
prescribed $Q$--CPU allocation under a common power-factor constraint. The
assessment is deliberately restricted to sinusoidal steady state and ideal
controlled current sources. It neither defines the inner control architecture
of a converter nor addresses current limits, dc-link management, switching
dynamics, or closed-loop stability.}

\rev{Independently of this compensation assessment, the CVP complements the
usual symmetrical-component viewpoint. Both phase and symmetrical-component
coordinates provide the conventional complex-power term $P+jQ$. However, the
separate sequence-power terms do not by themselves provide the cross-phase
vector constituent introduced in this work. The CVP makes this constituent
explicit through $\mathbf{D}$ in phase coordinates and through its transformed
representation in symmetrical-component coordinates, while preserving its norm
under the power-invariant Fortescue transformation.}

\rev{The same cross-phase structure also appears in the instantaneous
domain.} The vector $\mathbf{D}$ determines the amplitudes and phases of the associated $2\omega$
oscillatory components, and its norm relates to their rms magnitude. This is
consistent with classical observations that products of sinusoidal quantities
generate double-frequency terms~\cite{Steinmetz1899}. The present formulation
expresses these phase relationships through conventional complex arithmetic,
without requiring additional algebraic machinery.

Another point made explicit by the CVP is the role of the voltage reference in
apparent-power representations. The algebraic values and orientation of
$\mathbf{D}$ depend on the selected
voltage reference. Under the power-invariant Fortescue transformation,
however, its norm is preserved. Thus, the internal structure of the
representation is reference sensitive, while the associated apparent-power
metric remains well defined, in agreement with previous discussions on voltage
reference selection~\cite{Willems2003,Leon2020}. 

\rev{This reference issue is particularly relevant in three-phase four-wire
systems.} The equivalent-coordinate construction
introduced in Section~III provides a consistent way of working within a
three-coordinate setting without losing the effective apparent-power norm. This
allows the CVP to remain compatible with IEEE~1459 in the sinusoidal
steady-state regime while preserving the same multicomponent structure used in
the three-wire case.

\rev{From a computational viewpoint, the evaluation of the CVP, including
the equivalent-coordinate construction required for three-phase four-wire
systems, has a low computational burden once the sinusoidal steady-state
terminal phasors are available. The artificial-neutral displacement is
obtained from a closed-form linear combination of the measured phase voltages,
and the remaining operations consist of scalar multiplications, vector
products, and fixed $(3\times3)$ transformations. The CVP evaluation itself
requires no iterative optimization or network re-solution. Consequently, its
additional processing cost is small compared with the phasor-estimation stage
normally required in digital power-quality instruments or controllers. The
overall update rate, latency, and accuracy are therefore determined primarily
by the adopted phasor-estimation method and its observation window.
Measurement noise and phasor-estimation errors propagate algebraically to the
computed CVP quantities, including $\mathbf{D}$, as they do to conventional
phasor-domain power quantities; a detailed uncertainty assessment is outside
the scope of the present study. The iterative reference update used in the
compensation assessment arises instead from the network-dependent change in
terminal voltage produced by the finite feeder impedance and is not part of
the CVP evaluation.}

\rev{If fundamental-frequency phasors are estimated from mildly distorted terminal waveforms, the CVP norm identity remains algebraically valid for those estimated components. Harmonic distortion does not modify the identity itself, but it may affect the numerical values of $P$, $Q$, and $\mathbf{D}$ through the accuracy of the phasor-estimation process. The resulting quantities characterize only the fundamental-frequency voltage--current relation and do not include harmonic contributions, cross-frequency terms, or the total apparent power of the nonsinusoidal waveforms.}

\rev{This observation and the preceding implementation considerations do not alter the domain of the present formulation. The underlying network-analysis method may include detailed component models, mutual coupling, frequency-dependent elements, power-electronic devices, or nonlinear components. However, the resulting terminal operating state can be represented by the present CVP only when both the terminal voltages and currents satisfy the sinusoidal steady-state assumption. The CVP does not model switching processes, transient dynamics, harmonic distortion, or the total apparent power of nonsinusoidal waveforms. Higher-phase-order extensions are likewise not considered in the present work.}

\paragraph*{Relation to IEEE~1459 unbalance power}

\rev{Within this sinusoidal steady-state scope, the proposed decomposition can
be compared directly with the effective-apparent-power structure of
IEEE~1459.} The standard defines
$S_e^2 = (S^{+})^2 + S_u^2$. In the CVP formulation,
\[
\|\mathbf{S}\|^2 = P^2 + Q^2 + \|\mathbf{D}\|^2,
\qquad
\|\mathbf{S}\| = S_e.
\]
The two decompositions are therefore consistent at the level of the effective
apparent-power norm, but they differ in structure. IEEE~1459 defines the
unbalance power $S_u$ with respect to the positive-sequence apparent power,
thereby aggregating different effects of asymmetry into a scalar quantity. In
contrast, the CVP separates the total phase-domain interaction into an
intraphase contribution, represented by $P+jQ$, and an explicitly interphase
contribution, represented by $\mathbf{D}$.

Accordingly, $S_u$ does not in general coincide with $\|\mathbf{D}\|$. Rather
than replacing IEEE~1459 quantities, the CVP complements them by making explicit
the cross-phase contribution that remains implicit in scalar apparent-power
descriptions.

Importantly, the CVP does not postulate a new energetic carrier or an additional
physical kind of power. The cross-product term should be interpreted as a
structural descriptor of how the voltage--current relation is distributed across
phases under unbalanced conditions. In this sense, the CVP does not redefine the
physical mechanisms of energy transfer, but makes explicit a component of the
phase relations that is not displayed by scalar apparent-power
representations.

\section{Conclusions}

This work has introduced the Complex-Vector Power as a phasor-based formulation for sinusoidal unbalanced three-phase systems. \rev{By complementing the conventional complex-power scalar with a cross-phase unbalance vector}, the formulation makes the antisymmetric voltage--current relation explicit while preserving the apparent-power norm.

The CVP norm follows from the complex Lagrange identity and is preserved under the power-invariant Fortescue transformation. The formulation has also been extended to three-phase four-wire systems through equivalent coordinates that reproduce the effective rms voltage, current, and apparent-power quantities associated with IEEE~1459.

The CPU vector is not interpreted as an additional physical form of power. It is a structural descriptor that complements $P$ and $Q$ by identifying a cross-phase contribution to the apparent-power norm. This distinction becomes operationally relevant when power-factor degradation cannot be attributed to reactive power alone.

\rev{Finally, a steady-state shunt-compensation assessment shows that the
CPU descriptor can be incorporated into current references for selective CPU
reduction or prescribed $Q$--CPU allocation. This proof of concept supports
the engineering relevance of the CVP, without constituting the design or
validation of a complete converter-control system.}

\appendices
\section{Lagrange's Identity in Complex Form}

Let $\mathbf{V},\mathbf{I}\in\mathbb{C}^{n}$ be complex vectors endowed with the Euclidean norm $\|\cdot\|$. Lagrange's identity in complex form reads
\begin{equation}
\|\mathbf{V}\|^{2}\,\|\mathbf{I}\|^{2}
=
\left(\sum_{j=1}^{n}|\bar V_{j}|^{2}\right)
\left(\sum_{k=1}^{n}|\bar I_{k}|^{2}\right).
\label{eq:lagrange_start}
\end{equation}

This product can be decomposed as
\begin{equation}
\|\mathbf{V}\|^{2}\,\|\mathbf{I}\|^{2}
=
\left|
\sum_{j=1}^{n}\bar V_{j}\bar I_{j}^{*}
\right|^{2}
+
\sum_{1\leq j<k\leq n}
\left|
\bar V_{j}\bar I_{k}-\bar V_{k}\bar I_{j}
\right|^{2}.
\label{eq:lagrange_main}
\end{equation}

To see this, note that
\begin{equation}
\begin{split}
\sum_{1\leq j<k\leq n}
\left|
\bar V_{j}\bar I_{k}-\bar V_{k}\bar I_{j}
\right|^{2}
=
\frac{1}{2}
\sum_{j=1}^{n}\sum_{k=1}^{n}
\left|
\bar V_{j}\bar I_{k}-\bar V_{k}\bar I_{j}
\right|^{2} \\
=
\frac{1}{2}
\sum_{j=1}^{n}\sum_{k=1}^{n}
\Big(
|\bar V_{j}|^{2}|\bar I_{k}|^{2}
+
|\bar V_{k}|^{2}|\bar I_{j}|^{2}
-
2\,\Re\{\bar V_{j}\bar I_{j}^{*}\bar V_{k}^{*}\bar I_{k}\}
\Big).
\end{split}
\label{eq:lagrange_mid}
\end{equation}

Rearranging terms yields
\begin{align}
\sum_{1\leq j<k\leq n}
\left|
\bar V_{j}\bar I_{k}-\bar V_{k}\bar I_{j}
\right|^{2}
&=
\left(\sum_{j=1}^{n}|\bar V_{j}|^{2}\right)
\left(\sum_{k=1}^{n}|\bar I_{k}|^{2}\right) \nonumber\\
-
\Re\!\left\{
\sum_{j=1}^{n}\sum_{k=1}^{n}
\bar V_{j}\bar I_{j}^{*}
(\bar V_{k}\bar I_{k}^{*})^{*}
\right\}. \label{eq:lagrange_rearr}
\end{align}

Finally,
\begin{equation}
\Re\!\left\{
\sum_{j=1}^{n}\sum_{k=1}^{n}
\bar V_{j}\bar I_{j}^{*}
(\bar V_{k}\bar I_{k}^{*})^{*}
\right\}
=
\left|
\sum_{j=1}^{n}\bar V_{j}\bar I_{j}^{*}
\right|^{2}.
\label{eq:lagrange_final}
\end{equation}

Substitution of~\eqref{eq:lagrange_final} into~\eqref{eq:lagrange_rearr} yields~\eqref{eq:lagrange_main}.

This identity provides the mathematical basis for decomposing the norm of the
Complex-Vector Power into dot- and cross-product contributions.

\section{Cross Product under the Fortescue Transformation}

In symmetrical-component coordinates, the cross product is
\begin{equation}
\mathbf{V}^{\pm0}\times\mathbf{I}^{\pm0}
=
\begin{bmatrix}
\bar V^{-}\bar I^{0}-\bar V^{0}\bar I^{-}\\
\bar V^{0}\bar I^{+}-\bar V^{+}\bar I^{0}\\
\bar V^{+}\bar I^{-}-\bar V^{-}\bar I^{+}
\end{bmatrix}.
\end{equation}

Expressing these components in terms of phase-domain variables and simplifying with $1+a+a^{2}=0$ and $a^{3}=1$ yields
\begin{equation}
\mathbf{V}^{\pm0}\times\mathbf{I}^{\pm0}
=
\frac{1}{3}(a^{2}-a)
\begin{bmatrix}
1 & a^{2} & a\\
1 & a & a^{2}\\
1 & 1 & 1
\end{bmatrix}
(\mathbf{V}\times\mathbf{I}).
\end{equation}

Since
\begin{equation}
\det(\mathbf{A})=\frac{a^{2}-a}{\sqrt{3}}=-j,
\qquad
\mathbf{A}^{*}=\frac{1}{\sqrt{3}}
\begin{bmatrix}
1 & a^{2} & a\\
1 & a & a^{2}\\
1 & 1 & 1
\end{bmatrix},
\end{equation}
the transformation can be written compactly as
\begin{equation}
\mathbf{V}^{\pm0}\times\mathbf{I}^{\pm0}
=
\det(\mathbf{A})\,\mathbf{A}^{*}(\mathbf{V}\times\mathbf{I}).
\end{equation}

\vfill

\end{document}